\documentclass{svmult}

\usepackage{amsmath}
\usepackage{amssymb}
\usepackage{makeidx}         
\usepackage{graphicx}        
\usepackage{multicol}        

\makeindex             

\newcommand\beq{\begin{equation}}
\newcommand\eeq{\end{equation}}
\newcommand\beqa{\begin{eqnarray}}
\newcommand\eeqa{\end{eqnarray}}

\def\z{\zeta^*}

\begin{document}

\title*{Gravity-driven Poiseuille Flow in Dilute Gases. Elastic and Inelastic Collisions}
\titlerunning{Gravity-driven Poiseuille Flow in Dilute Gases}
\author{Andr\'es Santos\inst{1}\and
Mohamed Tij\inst{2}}
\institute{Departamento de F\'{\i}sica, Universidad de Extremadura,
E--06071 Badajoz, Spain \texttt{andres@unex.es} \and D\'epartment de
Physique, Universit\'e Moulay Isma\"{\i}l, Mekn\`es, Morocco
\texttt{mtij@fsmek.ac.ma}}
%
%
\maketitle

An overview of the hydrodynamic profiles derived by kinetic theory
tools (Boltzmann equation and kinetic models) for the gravity-driven
Poiseuille flow with  particles colliding either elastically (planar
and cylindrical geometries) or inelastically (planar geometry) is
presented. The results obtained through second order in the gravity
acceleration show that the Navier--Stokes predictions are
qualitatively incorrect over distances of the order of the mean free
path, especially in the case of the temperature profile.

\section{Introduction\label{sec0}}
 In the last decade, the Poiseuille flow generated
 by the action of a uniform longitudinal body force $m\vec{g}$
(e.g., gravity), rather than by a longitudinal pressure gradient,
has received much attention from  computational
\cite{TTE97,MBG97,RC98} and  theoretical
\cite{RC98,AS92,TS94,TSS98,UG99,HM99,TS01,ATN02,STS03,TS04} points
of view. This interest has been fueled by the fact that the
gravity-driven Poiseuille flow provides a nice example illustrating
the limitations of   the classical Navier--Stokes (NS) description,
even in the bulk domain (i.e., far away from the boundary layers),
over distances of the order of the mean free path. In contrast to
what happens with the structure of a plane shock wave, where the NS
discrepancies are essentially quantitative and are widely accounted
for  by the Burnett description~\cite{MHGS98}, here the main
failures of the NS predictions are qualitative and remain in the
Burnett theory.

Kinetic theory analyses of the  gravity-driven plane Poiseuille flow
based on an expansion in powers of the gravity acceleration $g$
\cite{TS94,TSS98},  on Grad's moment method \cite{RC98,HM99}, or on
an expansion in powers of the Knudsen number~\cite{ATN02}, show that
to second order in $g$ the temperature profile includes a positive
quadratic term  in addition to the negative quartic term predicted
by the NS (and Burnett) hydrodynamic equations. As a consequence of
this extra term, the temperature  does not present a flat maximum at
the middle of the channel but instead exhibits a \textit{bimodal}
shape with a local minimum
 surrounded by two symmetric maxima at a
distance of a few mean free paths.  The Fourier law is dramatically
violated since in the slab enclosed by the two maxima the transverse
component of the heat flux is parallel (rather than anti-parallel)
to the thermal gradient.  The kinetic theory prediction of a bimodal
temperature profile in a slab has been confirmed by computer
simulations \cite{MBG97,RC98,UG99}.
 A similar behavior occurs in the case of the Poiseuille flow in a
pipe \cite{TS01,STS03}.

For conventional gases under terrestrial conditions  the interest of
the Poiseuille flow induced by gravity is rather academic. However,
this is not necessarily so when dealing with a ``granular''
gas~\cite{C90}, i.e., a collection of a large number of discrete
solid particles (or grains) in a fluidized state such that each
particle moves freely and independently of the rest, except for the
occurrence of \textit{inelastic} binary collisions.
 A kinetic theory
study of a gas of inelastic hard spheres in a slab under the action
of a longitudinal force and excited by a white noise ``heating'' has
recently been undertaken~\cite{TS04}.

The aim of this paper is to provide an overview of the hydrodynamic
profiles derived by kinetic theory tools (Boltzmann equation and
kinetic models) for the gravity-driven Poiseuille flow in the cases
of  particles colliding either elastically (planar and cylindrical
geometries) or inelastically (planar geometry). The problem is
presented in Sect.~\ref{sec1} and the NS solution to order $g^2$ is
given in Sect.~\ref{sec2}. The kinetic theory description for
elastic and inelastic collisions is provided in Sects.\ \ref{sec3}
and \ref{sec4}, respectively. Finally, the main conclusions  are
briefly presented in Sect.~\ref{sec5}.

%
%
%

\section{Statement of the Problem\label{sec1}}
Let us consider a monodisperse dilute gas of identical spherical
particles of mass $m$. The interaction among particles is assumed to
reduce to binary collisions which can be either elastic or
inelastic. In the first case, the (local) kinetic energy per unit
volume  is preserved by collisions. On the other hand, if the
collisions are inelastic, energy is dissipated into the internal
degrees of freedom; we will assume that this \textit{internal}
energy sink is counterbalanced by an \textit{external} energy source
produced by some sort of driving (e.g., boundary vibrations,
external thermostats, \ldots). In addition, the system is under the
influence of a constant gravitational field characterized by the
acceleration vector $\vec{g}$. Under the above conditions, the
macroscopic balance equations for the local densities of mass,
momentum, and energy read
\beq
D_t n+n\nabla\cdot \vec{u}=0\;,\quad
D_t\vec{u}+\frac{1}{mn}\nabla\cdot\tens{P}=\vec{g}\;,
\label{b7}
\eeq
\beq
D_tT+\frac{2}{3n}\left(\nabla\cdot\vec{q}+\tens{P}:\nabla
\vec{u}\right)=-(\zeta-\gamma)T\;.
\label{b9}
\eeq
In these equations, $D_t\equiv\partial_t+\vec{u}\cdot\nabla$ is the
material time derivative, $n$ is the number density, $\vec{u}$ is
the mean velocity, $T$ is the temperature,\footnote{We have set the
Boltzmann constant equal to 1, so that $T$
 (which is usually termed ``granular temperature'' in the case of
 inelastic collisions) has dimensions of an energy.} $\tens{P}$ is
 the pressure or stress tensor, and $\vec{q}$ is the heat flux.
The trace of the pressure tensor defines the hydrostatic pressure
$p=\frac{1}{3}\text{Tr}\,\tens{P}$, which in a dilute gas is related
to the density and the temperature through the ideal gas equation of
state $p=nT$. In (\ref{b9}),  $\zeta\geq 0$ is the (local)
``cooling'' rate associated with the collisional inelasticity and
$\gamma\geq 0$ is the (local) ``heating'' rate associated with the
external driving. If the collisions are elastic, the cooling rate
vanishes ($\zeta=0$) and  no energy needs to be externally injected
into the system ($\gamma=0$). On the other hand, in the case of
inelastic collisions one has $\zeta>0$ and we will assume that the
external source of energy compensates \textit{locally} for the
collisional cooling, i.e., $\gamma=\zeta$.

\begin{figure}[tbp]
 \centering
\includegraphics[height=4cm]{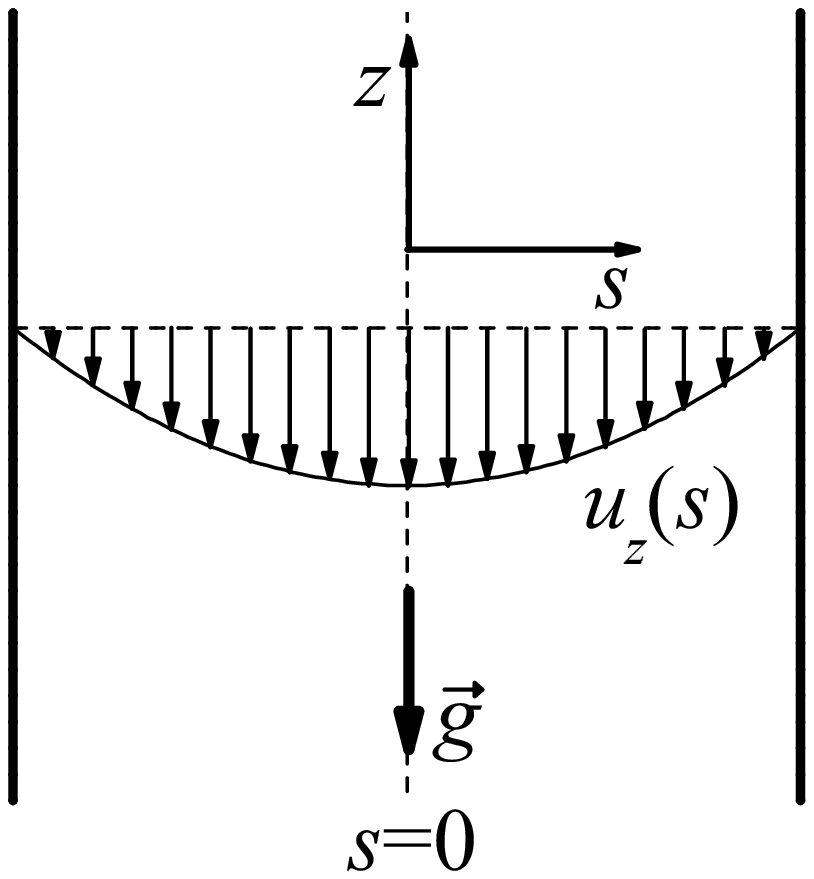}
\caption{Sketch of the longitudinal section of the Poiseuille flow
induced by a gravitational force. In the planar geometry, the
boundaries are infinite parallel plates and the variable $s$
represents the Cartesian coordinate $y$. In the cylindrical
geometry, the boundary is an infinite cylindrical surface and the
variable $s$ represents the radial coordinate $r$
\label{sketch}}
\end{figure}

 Now we particularize to the gravity-driven Poiseuille flow. In the
 planar geometry, the gas is enclosed between two infinite parallel
 plates normal to the $y$-axis and the acceleration of gravity
 $\vec{g}=-g
\vec{\widehat{z}}$ is applied along a direction
 $\vec{\widehat{z}}$ parallel to the plates. A longitudinal
 section of the system is sketched in Fig.~\ref{sketch}, where the
 variable $s$ refers to the coordinate $y$.
In the stationary laminar state the macroscopic variables depend on
the variable $y$ only, so the balance equations (\ref{b7}) and
(\ref{b9}) yield
\beq
\frac{\partial P_{yy}}{\partial y}=0\;,\quad  \frac{\partial
P_{yz}}{\partial y}=-\rho  g\;, \quad P_{yz}\frac{\partial
u_z}{\partial y}+ \frac{\partial q_{y}}{\partial y}=0\;,
\label{3}
\eeq
where $\rho=mn$ is the mass density and use has been made of the
assumptions $\zeta=\gamma=0$ (elastic collisions) or
$\zeta=\gamma>0$ (inelastic collisions). Equations (\ref{3}) imply
that the normal stress $P_{yy}$ is uniform across the system. Note
that neither the other two normal stresses $P_{xx}$ and $P_{zz}$ nor
the longitudinal component $q_z$ of the heat flux appear explicitly
in the balance equations (\ref{3}).

The Poiseuille flow can also be defined in a cylindrical geometry.
In that case, the gas is inside an infinitely long straight tube of
circular cross section, parallel to the $z$-axis. Figure
\ref{sketch} also represents a longitudinal section of the system,
except that now the variable $s$ refers to the radial coordinate
$r=\sqrt{x^2+y^2}$. Assuming again a stationary laminar flow, the
only relevant spatial variable is $r$. Expressing the pressure
tensor and the heat flux in cylindrical coordinates,  the balance
equations (\ref{b7}) and (\ref{b9}) reduce to \cite{TS01,STS03}
\beq
\frac{\partial}{\partial r}\left(r P_{rr}\right)=P_{\phi\phi}\;,
\quad \frac{1}{r}\frac{\partial}{\partial r}\left(r
P_{rz}\right)=-\rho g\;, \quad P_{rz} r\frac{\partial u_z}{\partial
r}+\frac{\partial}{\partial r}\left(r q_r\right)=0\;.
\label{2.15a}
\eeq

Our aim is to get the hydrodynamic profiles $p(s)=n(s)T(s)$,
$u_z(s)$, and $T(s)$ (where $s=y$ and $s=r$ for the planar and
cylindrical geometries, respectively) in the \textit{bulk} domain,
i.e., around $s=0$, by a perturbation expansion through second order
in $g$.

\section{Navier--Stokes Theory\label{sec2}}
Equations (\ref{3}) (planar case) and (\ref{2.15a}) (cylindrical
case) do not constitute a closed set of equations. In a purely
hydrodynamic approach, they must be complemented by constitutive
relations expressing the fluxes $\tens{P}$ and $\vec{q}$ in terms of
the hydrodynamic fields and their gradients.

If the gravity field were switched off ($g=0$), the physical
solution to (\ref{3})  and (\ref{2.15a})   would correspond to an
isotropic and uniform state, i.e., $\nabla p=\nabla T=\nabla
\vec{u}=\vec{0}$, $\tens{P}=p\,\tens{I}$ (where $\tens{I}$ is the
unit tensor), and $\vec{q}=\vec{0}$. This is not but the equilibrium
state in the case of elastic collisions and a uniformly heated
non-equilibrium steady state in the case of inelastic collisions. On
the other hand, the existence of the constant field $g$, no matter
how small it is, induces gradients along the directions normal to
the boundaries. It seems then quite natural that a Navier--Stokes
(NS) description might be appropriate, at least for asymptotically
small values of $g$. As will be seen, this expectation turns out not
to hold true.

In a monodisperse gas, the NS equations follow from Newton's
viscosity law and Fourier's heat conduction law,
\beq
\tens{P}=p\,\tens{I}-\eta\left[\nabla \vec{u}+\left(\nabla
\vec{u}\right)^\dag-\frac{2}{3}\left(\nabla\cdot\vec{u}\right)\tens{I}\right]\;,
\quad \vec{q}=-\kappa\nabla T-\mu\nabla n\;,
\label{n2.1}
\eeq
where the dagger denotes the transpose, $\eta$ is the shear
viscosity, $\kappa$ is the thermal conductivity, and $\mu$ is a
coefficient that vanishes if the collisions are elastic but is in
general non-zero in granular gases. The expressions of the transport
coefficients $\eta$ and $\kappa$ in terms of the temperature and the
mechanical properties of the particles are well known in the case of
elastic systems~\cite{CC70}. Moreover, $\eta$, $\kappa$, and $\mu$
have been derived for inelastic hard spheres \cite{BDKS98,GM01} and
inelastic Maxwell particles~\cite{S03} in the free cooling case as
well as under heating.

Since in the Poiseuille flow the gradients are normal to the flow
direction, it follows from (\ref{n2.1}) that the three normal
stresses are identical. In Cartesian and cylindrical coordinates,
\beq
P_{xx}=P_{yy}=P_{zz}=p\;,\quad P_{\phi\phi}=P_{rr}=P_{zz}=p\;,
\label{n2.3}
\eeq
respectively. Equations (\ref{3}) or (\ref{2.15a}) then imply that
$p=\text{const}$, so that the (modified) Fourier law in (\ref{n2.1})
becomes
\beq
\vec{q}=-\kappa'\nabla T\;,
\label{n2.6}
\eeq
where we have called $\kappa'\equiv \kappa-n\mu/T$. Moreover, there
does not exist a longitudinal component of the heat flux.

When the NS constitutive relations (\ref{n2.1})  are used in the
balance equations  (\ref{3}) for the planar case, one gets the
closed set of equations
\beq
\frac{\partial}{\partial y}\eta\frac{\partial u_z}{\partial
y}=\frac{mpg}{T}\;,\quad \frac{\partial}{\partial
y}\kappa'\frac{\partial T}{\partial y}=- \eta\left(\frac{\partial
u_z}{\partial y}\right)^2\;.
\label{2.7}
\eeq
Equations (\ref{2.7}) give a parabolic-like velocity profile, that
is characteristic of the Poiseuille flow, and a quartic-like shape
for the temperature profile. Strictly speaking, these NS profiles
are more complicated than just polynomials due to the temperature
dependence of the transport coefficients. The knowledge of such a
dependence is needed in order to get the NS hydrodynamic profiles
from (\ref{2.7}). However, if one is interested in those profiles to
second order in $g$, the precise dependence of $\eta$ and $\kappa'$
on $T$ is not needed and one can replace $\eta\to\eta_0$ and
$\kappa'\to\kappa_0'$.\footnote{Henceforth the subscript 0 attached
to a quantity will denote the value of that quantity at the mid
plane $y=0$  or at the symmetry axis $r=0$.} More explicitly, the
solution to
 (\ref{2.7}) through order $g^2$ is
\beq
u_z(y)=u_0+\frac{\rho_0 g}{2\eta_0}{y}^2+ { O}(g^3)\;, \quad
p(y)=p_0\;,
\label{2.11.1}
\eeq
\beq
T(y)=T_0\left(1-\frac{\rho_0^2
g^2}{12\eta_0\kappa_0'T_0}{y}^4\right) +{O}(g^4)\;.
\label{2.12.1}
\eeq
 The shear
stress and  the heat flux are
\beq
P_{yz}(y)=-\rho_0 g{y}\left(1+\frac{\rho_0^2
g^2}{60\eta_0\kappa_0'T_0}{y}^4\right)+{O}(g^5)\;,
\label{2.11.3}
\eeq
\beq
q_{y}(y)=\frac{\rho_0^2 g^2}{3\eta_0}{y}^3+{O}(g^4)\;,\quad
q_z(y)=0\;,
\label{2.11.4}
\eeq
respectively. Note that the knowledge of the temperature profile to
$O(g^2)$ allows us to get $P_{xy}$ to $O(g^3)$ through (\ref{3}).
Equation (\ref{2.12.1}) shows that the NS solution predicts that the
temperature has a maximum at $y=0$.

In the cylindrical case, insertion of (\ref{n2.1}) into
(\ref{2.15a})  yields
\beq
r^{-1}\frac{\partial}{\partial r}\left(\eta r\frac{\partial
u_z}{\partial r}\right)=\frac{mpg}{T}\;,\quad
r^{-1}\frac{\partial}{\partial r}\left(\kappa' r\frac{\partial
T}{\partial r}\right)=-\eta\left(\frac{\partial u_z}{\partial
r}\right)^2\;.
\label{2.24}
\eeq
Again, these two equations can be easily solved to second order in
$g$:
\beq
u_z(r)=u_0+\frac{\rho_0 g}{4\eta_0}{r}^2+\mathcal{O}(g^3)\;,\quad
p(r)=p_0\;,
\label{2.25}
\eeq
\beq
T(r)=T_0\left(1-\frac{1}{64}\frac{\rho_0^2
g^2}{\eta_0\kappa_0'T_0}{r}^4\right) +\mathcal{O}(g^4)\;.
\label{2.26}
\eeq
 The corresponding expressions for the shear stress and the  heat  flux are
\beq
P_{rz}(r)=-\frac{\rho_0 g}{2}{r}\left(1+\frac{1}{192}\frac{\rho_0^2
g^2}{\eta_0\kappa_0'T_0}{r}^4\right)+\mathcal{O}(g^5)\;,
\label{2.26.1}
\eeq
\beq
q_{r}(r)=\frac{\rho_0^2 g^2}{16\eta_0}{r}^3+\mathcal{O}(g^4)\;,
\quad q_z(r)=0\;,
\label{2.26.2}
\eeq
respectively. Equation (\ref{2.26}) implies that the temperature is
maximum at $r=0$.
\section{Kinetic Theory Description. Elastic Collisions\label{sec3}}
In a kinetic theory approach the relevant information of the system
is conveyed by the one-particle velocity distribution function
$f(\vec{r},\vec{v},t)$. Its temporal evolution is governed by the
non-linear Boltzmann equation~\cite{CC70}, which in standard
notation reads
\beq
\label{ab8}
\frac{\partial}{\partial t}f+\vec{v}\cdot \nabla f+
\vec{g}\cdot\frac{\partial}{\partial\vec{v}}f= J[f,f]\;,
\eeq
where $J[f,f]$ is the collision operator for elastic collisions and
we have assumed that no external forcing exists, apart from gravity.
The influence of the interaction potential appears in $J[f,f]$
through the dependence of the collision rate on the relative
velocity  and the scattering angle. The hydrodynamic variables and
their fluxes are just velocity moments of the distribution function:
\begin{equation}
 \left\{n,n\vec{u},p,\tens{P},\vec{q}\right\} = { \int } \D \vec{v} \,
 \left\{1,\vec{v},\frac{m}{3}V^2,m\vec{V}\vec{V},\frac{m}{2}V^2\vec{V}\right\} f \;,
\label{16}
\end{equation}
where the  peculiar velocity $ \vec{V}=\vec{v}-\vec{u}$ has been
introduced as the velocity of a particle relative to the flow
velocity. The exact balance equations (\ref{b7}) and (\ref{b9}),
with $\zeta=\gamma=0$, follow from (\ref{ab8}) by taking velocity
moments.

\subsection{Planar Geometry}
In the case of the stationary Poiseuille flow in a channel, the
Boltzmann equation (\ref{ab8}) becomes
\begin{equation}
\left(v_y \frac{\partial }{\partial y} - g \frac{\partial }{\partial
v_z} \right)f= J \lbrack f, f \rbrack \;.
\label{22}
\end{equation}
Strictly speaking, (\ref{22}) must be supplemented by the
appropriate boundary conditions  describing the interaction of the
particles with the plates. However,  we are interested in the
\textit{bulk} region of the system, sufficiently far from the
boundaries. Therefore, we will look for the Hilbert-class or
\textit{normal} solution to (\ref{22}), namely a solution where all
the spatial dependence of the distribution function takes place
through a \textit{functional} dependence of $f$ on the hydrodynamic
fields $n$, $\vec{u}$, and $T$.

Since the hydrodynamic variables of the gas and the associated
fluxes are the first few moments of the distribution function $f$,
it is convenient to consider the hierarchy of moment equations
stemming from the Boltzmann equation (\ref{22}). A moment of an
arbitrary degree $k=k_1+k_2+k_3$ is defined as
\begin{equation}
  M_{k_1,k_2,k_3} (y;g)= { \int } \D\vec{v}\, V^{k_1}_{x} V^{k_2}_{y} V^{k_3}_{z} f(y,\vec{v};g)\;.
\label{23}
\end{equation}
Because of the symmetry properties of the problem, $M_{k_1,k_2,k_3}$
is a  function of $y$ of the same parity as $k_2$. Seen as a
function of $g$, $M_{k_1,k_2,k_3}$ has the same parity as $k_3$.
Finally, $M_{k_1,k_2,k_3}=0$ if $k_1=\text{odd}$. {}From (\ref{22})
one gets the hierarchy
\beq
\label{b3}
\frac{\partial}{\partial y}M_{k_1,k_2+1,k_3}+k_3\left(\frac{\partial
u_z}{\partial y} M_{k_1,k_2+1,k_3-1} +gM_{k_1,k_2,k_3-1}\right)=
J_{k_1,k_2,k_3}\;,
\eeq
where $J_{k_1,k_2,k_3}$ is a collisional moment. In general, it
involves all the velocity moments of the distribution, including
those of a higher degree, and its explicit expression in terms of
those moments is unknown. An important exception is provided by the
Maxwell interaction potential, in which case the collision rate is
independent of the velocity and, as a consequence, $J_{k_1,k_2,k_3}$
can be expressed as a bilinear combination of moments of the same or
smaller degree.

Even in the case of Maxwell molecules, the hierarchy (\ref{b3})
couples moments of a certain degree $k$ to moments of degree
$k+1$.\footnote{In what follows, we will  use the roman boldface
$\vec{k}$ to denote the triad $\{k_1,k_2,k_3\}$ and the italic
lightface $k$ to denote the sum $k_1+k_2+k_3$.} However, the problem
can be solved by a recursive scheme~\cite{TSS98}. First, the moments
are expanded in powers of $g$ around a reference equilibrium state
parameterized by $u_0$, $p_0$, and $T_0$:
\beq
u_z(y;g)=u_0+\sum_{\ell=1}^\infty u_z^{(\ell)}(y) g^\ell\;,\quad
p(y;g)=p_0+\sum_{\ell=2}^\infty p^{(\ell)}(y) g^\ell\;,
\label{n4.1}
\eeq
\beq
T(y;g)=T_0+\sum_{\ell=2}^\infty T^{(\ell)}(y) g^\ell\;,\quad
M_{\vec{k}}(y;g)=M_{\vec{k}}^{(0)}+\sum_{\ell=1}^\infty
M_{\vec{k}}^{(\ell)}(y) g^\ell\;.
\label{n4.4}
\eeq
 Due
to symmetry reasons, $u_z^{(\ell)}(y)=0$ if $\ell=\text{even}$,
$p^{(\ell)}(y)=T^{(\ell)}(y)=0$ if $\ell=\text{odd}$, and
$M_{\vec{k}}^{(\ell)}(y)=0$ if $k_3+\ell=\text{odd}$. The second
step consists of the ansatz that the coefficients have just a
polynomial dependence on $y$, namely
\beq
u_z^{(\ell)}(y)=\sum_{j=1}^\ell u_z^{(\ell,2j)}y^{2j}\;,\quad
p^{(\ell)}(y)=\sum_{j=2}^\ell p^{(\ell,2j)}y^{2j}\;,
\label{n4.5}
\eeq
\beq
T^{(\ell)}(y)=\sum_{j=2}^\ell T^{(\ell,2j)}y^{2j}\;,\quad
M_{\vec{k}}^{(\ell)}(y)=\sum_{j=0}^{N_\ell}M_{\vec{k}}^{(\ell,j)}y^j\;,
\label{n4.6}
\eeq
where $N_1=1$ and $N_\ell=2\ell$ for $\ell\geq 2$. Symmetry implies
that $M_{\vec{k}}^{(\ell,j)}=0$ if $k_2+j=\text{odd}$. The numerical
coefficients $u_z^{(\ell,2j)}$, $p^{(\ell,2j)}$, $T^{(\ell,2j)}$,
and $M_{\vec{k}}^{(\ell,j)}$ are  determined recursively by
inserting (\ref{n4.1})--(\ref{n4.6}) into (\ref{b3}) and equating
the coefficients of the same powers in $g$ and $y$  in both sides.
This yields a hierarchy of \textit{linear} equations for the
unknowns. This rather cumbersome scheme has been solved through
order $g^2$ in \cite{TSS98}. The results for the hydrodynamic
profiles and the fluxes are
\beq
u_z(y)=u_0+\frac{\rho_0 g}{2\eta_0}{y}^2+O(g^3)\;,\quad
p(y)=p_0\left[1+C_p\left(\frac{mg}{T_0}\right)^2y^2\right]+ {
O}(g^4)\;,
\label{3.9}
\eeq
\beq
T(y)=T_0\left[1-\frac{\rho_0^2 g^2}{12\eta_0\kappa_0T_0}{y}^4+
C_T\left(\frac{mg}{T_0}\right)^2y^2\right] +{ O}(g^4)\;,
\label{3.11}
\eeq
\beq
P_{zz}(y)=p_0\left[1+\frac{7}{3}C_p
\left(\frac{mg}{T_0}\right)^2y^2+C_\varpi\frac{\rho_0\eta_0^2g^2}{p_0^3}\right]+
{O}(g^4)\;,
\label{3.13}
\eeq
\beq
P_{yy}=p_0\left(1-C_\varpi'\frac{\rho_0\eta_0^2g^2}{p_0^3}\right)+
{\cal O}(g^4)\;,
\label{3.14}
\eeq
\beq
P_{yz}(y)=-\rho_0gy\left[1+\frac{\rho_0^2
g^2}{60\eta_0\kappa_0T_0}{y}^4+\frac{C_p-C_T}{3}
\left(\frac{mg}{T_0}\right)^2y^2\right]+ { O}(g^5)\;,
\label{3.16}
\eeq
\beq
q_y(y)=\frac{\rho_0^2g^2}{3\eta_0}y^3+{\cal O}(g^4)\;,\quad
q_z(y)=C_q{m g \kappa_0} +{\cal O}(g^3)\;.
\label{3.18}
\eeq
The numerical values of the coefficients $C_p$, $C_T$, $C_\varpi$,
$C_\varpi'$, and $C_q$ are displayed in Table \ref{table1}.
Comparison of (\ref{3.9})--(\ref{3.18}) with the NS predictions
(\ref{n2.3}) and (\ref{2.11.1})--(\ref{2.11.4}) shows that the NS
approach fails to account for the non-zero values of those five
coefficients. At a qualitative level, the main contrast appears for
the temperature profile. While, according to the NS equations, the
temperature has a maximum at the mid plane $y=0$, the Boltzmann
equation shows that, because of the extra quadratic term headed by
$C_T$, the temperature actually presents a local minimum $T_0$ at
$y=0$ surrounded by two symmetric maxima $T_{\text{max}}$ at $y=\pm
y_{\text{max}}$, where $y_{\text{max}}\equiv
\sqrt{6C_T\eta_0\kappa_0T_0}/p_0$. The relative height of the maxima
is $(T_\text{max}-T_0)/T_0= 2C_T(gy_{\text{max}}/v_0^2)^2$, where
$v_0=\sqrt{2T_0/m}$ is the thermal velocity at $y=0$. If one defines
an effective mean free path as $\lambda=16\sqrt{2\eta\kappa
T/15\pi}/5p$,\footnote{This definition guarantees that
$\lambda=(\sqrt{2}\pi n\sigma^2)^{-1}$ for hard spheres of diameter
$\sigma$.} one can write $y_{\text{max}}=(15\sqrt{5\pi
C_T}/16)\lambda_0\simeq 3.74\lambda_0$, $(T_\text{max}-T_0)/T_0=
(1125\pi C_T^2/128)(g\lambda_0/v_0^2)^2\simeq
28.46(g\lambda_0/v_0^2)^2$.
\begin{table}
\centering \caption{Numerical values of the coefficients $C_p$,
$C_T$, $C_\varpi$, $C_\varpi'$, and $C_q$, both in the planar and
the cylindrical geometries, according to the Navier--Stokes
description (NS), the Boltzmann equation for Maxwell molecules
(B-M), and the Bhatnagar--Gross--Krook kinetic model for any
interaction (BGK)}
\label{table1}       
%
%
\begin{tabular}{cccccccc}
\hline\noalign{\smallskip}
&\multicolumn{3}{c}{planar}&&\multicolumn{3}{c}{cylindrical}\\
\cline{2-4} \cline{6-8}\noalign{\smallskip}
coefficient & NS & B-M &BGK & &NS & B-M &BGK \\
\noalign{\smallskip}\hline\noalign{\smallskip}
$C_p$ & $0$ & $1.2$ &$1.2$ & &$0$ & $0.3$&$0.3$\\
$C_T$ & $0$ & $1.0153$ &$0.76$ && $0$ & $0.19429$&$0.14$\\
$C_\varpi$ & $0$ & $6.4777$ &$13.12$ && $0$ & $3.4776$&$7.36$\\
$C_\varpi'$ & $0$ & $6.2602$ &$12.24$ && $0$ & $1.7388$&$3.68$\\
$C_q$ & $0$ & $0.4$ &$0.4$ & &$0$ &
$0.4$&$0.4$\\\noalign{\smallskip}\hline
\end{tabular}
\end{table}

Equations (\ref{3.9})--(\ref{3.18}) are exact in the context of the
Boltzmann equation for Maxwell molecules. For more realistic
potentials (e.g., hard spheres) the hierarchy (\ref{b3}) cannot be
solved recursively, so that the problem must be addressed by means
of approximations. One theoretical possibility is to consider the
Chapman--Enskog expansion and retain as many terms as needed to get
the profiles through order $g^2$, but this is not very practical.
The coefficients $C_p$ and $C_q$ are already captured by the Burnett
description \cite{TSS98,UG99}, and so the Maxwell values
$C_p=\frac{6}{5}=1.2$ and $C_q=\frac{2}{5}=0.4$ can be expected to
be good approximations for other potentials. On the other hand, the
determination of $C_T$, $C_\varpi$, and $C_\varpi'$ requires the
consideration of super-Burnett  and super-super-Burnett
contributions~\cite{TSS98}. A second approach is Grad's moment
method~\cite{G49}.  The thirteen-moment approximation~\cite{RC98}
yields (\ref{3.9})--(\ref{3.18}) with $C_p=\frac{6}{5}$,
$C_q=\frac{2}{5}$, $C_T=\frac{14}{25}=0.56$, and
$C_\varpi=C_\varpi'=0$, irrespective of the interaction potential.
On the other hand, the nineteen-moment approximation~\cite{HM99}
gives $C_p=\frac{50998}{42025}\simeq 1.214$ and
$C_T=\frac{208518}{210125}\simeq 0.992$ for hard spheres, while it
predicts $C_p=\frac{6}{5}$ and $C_T=\frac{26}{25}=1.04$ for Maxwell
molecules. Comparison with the exact results $C_p=\frac{6}{5}$ and
$C_T=1.0153$ for Maxwell molecules suggests  that the
nineteen-moment approximation is rather accurate, although it
overestimates $C_T$ by a few percent.

An alternative route to get reasonable estimates with  relatively
much less effort than in the case of the Boltzmann equation consists
of using a kinetic model equation. The prototype kinetic model is
the one proposed by Bhatnagar, Gross, and Krook (BGK) and,
independently, by Welander and Kogan~\cite{BGK}. In this model, the
Boltzmann collision operator is replaced by a relaxation-time  term
towards the local equilibrium state:
\beq
J[f,f]\to -\nu\left(f-f_\ell\right)\;,\quad f_{\ell }=n\left(
\frac{m}{2\pi T} \right) ^{3/2}\exp \left( -{mV^2}/{2T}\right)\; .
\label{m23}
\end{equation}
The BGK model has proven to be quite reliable in non-Newtonian shear
flow problems~\cite{book}. The solution to the BGK equation for the
plane Poiseuille flow has been explicitly obtained through order
$g^5$~\cite{TS94}. The results strongly suggest that the series
expansion is only asymptotic, so that from a practical point of view
one can focus on the first few terms. The results agree with the
profiles (\ref{3.9})--(\ref{3.18}), except that the numerical values
of the coefficients $C_T$, $C_\varpi$, and $C_\varpi'$ differ from
those derived from the Boltzmann equation for Maxwell molecules, as
shown in Table \ref{table1}. The coefficient $C_T$ in the BGK model
is about 25\% smaller than in the Boltzmann equation, while
$C_\varpi$ and $C_\varpi'$ are about twice larger in the former than
in the latter. Interestingly enough, the BGK value
$C_T=\frac{19}{25}=0.76$ agrees quite well with Monte Carlo
simulations of the Boltzmann equation for hard spheres \cite{MBG97},
even though the nineteen-moment method predicts $C_T\lesssim 0.99$.
\subsection{Cylindrical Geometry}
In the cylindrical Poiseuille flow, the Boltzmann equation
(\ref{ab8}) becomes
\beq
\left[v_r\frac{\partial}{\partial r}+\frac{v_\phi}{r}\left(v_\phi
\frac{\partial }{\partial v_r}-v_r \frac{\partial }{\partial
v_\phi}\right)-g\frac{\partial}{\partial v_z}\right]f=J[f,f]\;.
\label{4.1}
\eeq
The algebra is now more involved than in the planar case, but still
the hierarchy of moment equations can be recursively solved for
Maxwell molecules through an expansion in powers of $g$. To second
order, the solution is~\cite{STS03}
\beq
u_z(r)=u_0+\frac{\rho_0 g}{4\eta_0}{r}^2+O(g^3)\;,\quad
p(r)=p_0\left[1+C_p\left(\frac{mg}{T_0}\right)^2r^2\right]+ {
O}(g^4)\;,
\label{c3.9}
\eeq
\beq
T(z)=T_0\left[1-\frac{\rho_0^2 g^2}{64\eta_0\kappa_0T_0}{r}^4+
C_T\left(\frac{mg}{T_0}\right)^2r^2\right] +{ O}(g^4)\;,
\label{c3.11}
\eeq
\beq
P_{zz}(y)=p_0\left[1+\frac{7}{3}C_p
\left(\frac{mg}{T_0}\right)^2r^2+C_\varpi\frac{\rho_0\eta_0^2g^2}{p_0^3}\right]+
{O}(g^4)\;,
\label{c3.13}
\eeq
\beq
P_{rr}(r)=p_0\left(1+\frac{1}{6}C_p
\left(\frac{mg}{T_0}\right)^2r^2-C_\varpi'\frac{\rho_0\eta_0^2g^2}{p_0^3}\right)+
{\cal O}(g^4)\;,
\label{c3.14}
\eeq
\beq
P_{rz}(r)=-\frac{\rho_0g}{2}r\left[1+\frac{\rho_0^2
g^2}{192\eta_0\kappa_0T_0}{r}^4+\frac{C_p-C_T}{2}
\left(\frac{mg}{T_0}\right)^2r^2\right]+ { O}(g^5)\;,
\label{c3.16}
\eeq
\beq
q_r(r)=\frac{\rho_0^2g^2}{16\eta_0}r^3+{\cal O}(g^4)\;, \quad
q_z(r)=C_q{m g \kappa_0} +{\cal O}(g^3)\;.
\label{c3.18}
\eeq
The numerical values of the coefficients $C_p$, $C_T$, $C_\varpi$,
$C_\varpi'$, and $C_q$ for this case are again displayed in Table
\ref{table1}. As happened in the planar case, the temperature
exhibits a non-monotonic profile: it has a local minimum  $T_0$ at
$r=0$ and its maximum value $T_{\text{max}}$ is reached at $r=
r_{\text{max}}$, where $r_{\text{max}}\equiv
\sqrt{32C_T\eta_0\kappa_0T_0}/p_0= (5\sqrt{15\pi
C_T}/4)\lambda_0\simeq 3.78\lambda_0$. The relative value of the
maximum is $(T_\text{max}-T_0)/T_0=
2C_T(gr_{\text{max}}/v_0^2)^2=(375\pi
C_T^2/8)(g\lambda_0/v_0^2)^2\simeq 5.56(g\lambda_0/v_0^2)^2$. It is
noteworthy that, expressed in the same units, $r_{\text{max}}$ is
practically equal to $y_{\text{max}}$, while $T_\text{max}-T_0$ is
about five times smaller in the cylindrical geometry than in the
planar geometry.

The BGK model has been solved for the cylindrical Poiseuille flow
through order $g^4$~\cite{TS01}, suggesting again an asymptotic
character of the expansion. The resulting profiles agree with
(\ref{c3.9})--(\ref{c3.18}), but with different values of $C_T$,
$C_\varpi$, and $C_\varpi'$ (see Table \ref{table1}). It is
interesting to note that the ratios between the BGK and the B-M
values for $C_T$, $C_\varpi$, and $C_\varpi'$ are almost the same in
the planar and the cylindrical geometries.

\begin{figure}[tbp]
 \centering
\includegraphics[height=4cm]{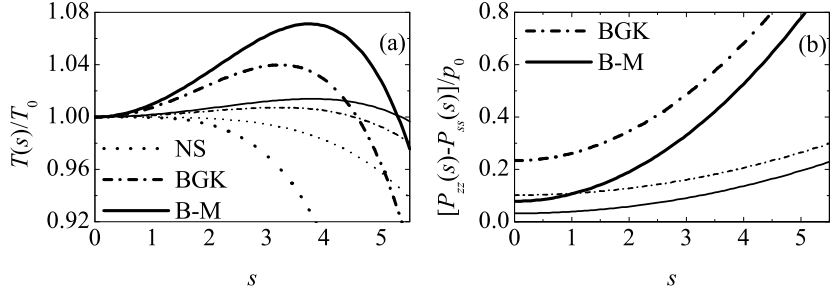}
\caption{Plot of (\textbf{a}) $T(s)/T_0$ and (\textbf{b})
$[P_{zz}(s)-P_{ss}(s)]/p_0$ for $g\lambda_0/v_0^2=0.05$ in the
planar ($s=y$, \textit{thick lines}) and cylindrical ($s=r$,
\textit{thin lines}) geometries, according to the NS equations
($\cdots$), the BGK model (-- $\cdot$ -- $\cdot$ -- ), and the
Boltzmann equation for Maxwell molecules (---)
\label{fig1}}
\end{figure}

Figure \ref{fig1} shows the profiles for the temperature  and for
the normal stress difference $P_{zz}-P_{ss}$ (where $P_{ss}=P_{yy}$
and $P_{ss}=P_{rr}$ in the planar and cylindrical geometries,
respectively) for $g\lambda_0/v_0^2=0.05$.

\section{Kinetic Theory Description. Inelastic Collisions\label{sec4}}
Now we consider the plane Poiseuille flow and assume that the
particles are smooth \textit{inelastic} hard spheres of diameter
$\sigma$ and coefficient of restitution $\alpha<1$. In the dilute
regime, the cooling rate  is approximately given by
\cite{BDKS98,BDS99}
\begin{equation}
\zeta(\alpha)=\nu \frac{5}{12} (1-\alpha ^{2})\;,\quad  \nu\equiv
\frac{16}{5}n\sigma ^{2}\left({\pi T}/{m} \right) ^{1/2}\;.
\label{n3.14}
\end{equation}
It is obvious that all the mathematical difficulties embodied in the
Boltzmann equation for (elastic) hard spheres are further increased
in the inelastic case. Therefore, in order to get explicit results
with a reasonable amount of effort, it seems convenient to extend
the BGK model (\ref{m23}) to the realm of inelastic collisions. The
simplest possibility is perhaps \cite{BDS99,SA04}
\beq
J[f,f]\to -\beta(\alpha)\nu (f-f_{\ell })+\frac{\zeta(\alpha)}{2}
\frac{\partial}{\partial\vec{v}}\cdot \left(\vec{V} f\right)\; ,
\label{m22}
\end{equation}
where $\beta(\alpha)$ is a dimensionless function of the coefficient
of restitution that can be freely chosen to optimize the agreement
with the Boltzmann description. A simple choice  is
$\beta(\alpha)=\frac{1}{2}(1+\alpha)$~\cite{SA04}. Here, however, we
will take $\beta(\alpha)=\frac{1}{6}(1+\alpha)(2+\alpha)$, which
makes the BGK shear viscosity  agree well with the Boltzmann one
\cite{BDS99,SA04}. Good quantitative agreement between the  kinetic
model (\ref{m22}) and the Boltzmann equation has been found for the
simple shear and Couette flows~\cite{BRM97}.

As said in Sect.~\ref{sec1}, an external energy input is needed to
compensate for the collisional dissipation in order to reach a
nonequilibrium steady state. In real experiments
 this is usually achieved by means of
boundary vibrations of small amplitude and high frequency. However,
this type of realistic heating through the boundaries is difficult
to deal with at a theoretical level due to  unavoidable boundary
effects. These difficulties are overcome by assuming a \textit{bulk}
heating mechanism acting on all the particles simultaneously. The
most commonly used type of bulk driving for inelastic particles
consists of a stochastic force in the form of Gaussian white noise,
which appears in the Boltzmann equation under the form of a
diffusion term in velocity space~\cite{vNE98}. In summary, our
kinetic equation to describe the stationary plane Poiseuille flow
reads~\cite{TS04}
\beq
\left(v_y\frac{\partial }{\partial y}-g\frac{\partial}{\partial
v_z}-\frac{\gamma T}{2m}\frac{\partial^2}{\partial \vec{v}^2}\right)
f= -\beta\nu (f-f_{\ell })+\frac{\zeta}{2}
\frac{\partial}{\partial\vec{v}}\cdot \left(\vec{V} f\right)\;,
\label{21}
\eeq
where the heating rate $\gamma$ is taken equal to $\zeta$, as
discussed in Sect. \ref{sec1}.

\begin{figure}[tbp]
 \centering
\includegraphics[height=4cm]{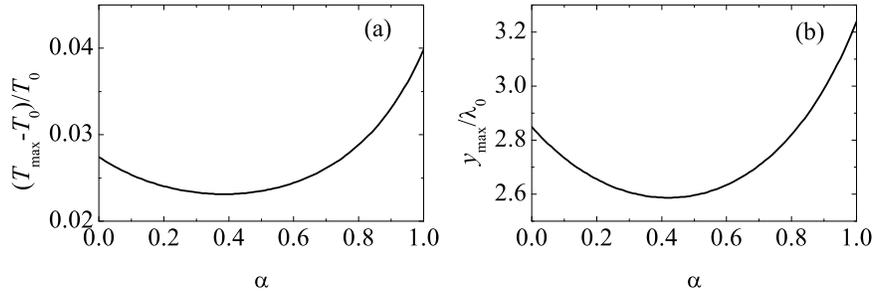}
\caption{Plot of (\textbf{a}) $(T_{\text{max}}-T_0)/T_0$ for
$g\lambda_0/v_0^2=0.05$ and (\textbf{b}) $y_{\text{max}}/\lambda_0$
as functions of the coefficient of restitution $\alpha$
\label{fig2}}
\end{figure}

The perturbative solution of (\ref{21}) is seen to agree with the
structure of (\ref{3.9})--(\ref{3.18}), except that now the NS
transport coefficients $\eta$ and $\kappa'=\kappa-n\mu/T$, as well
as $C_T$, $C_\varpi$,  and  $C_\varpi$, are functions of the
coefficient of restitution. As said above, the choice
$\beta=\frac{1}{6}(1+\alpha)(2+\alpha)$ guarantees that the BGK
shear viscosity $\eta$ agrees with the one derived from the
Boltzmann equation for inelastic hard spheres in the simplest Sonine
approximation. However, as already happens in the elastic case, the
(modified) thermal conductivity $\kappa'$ differs in both kinetic
equations. In order to circumvent this problem, once the BGK
solution is expressed in terms of $\eta$ and $\kappa'$, we will take
for those coefficients the expressions obtained from the Boltzmann
equation. In the case of a granular gas  heated by the white noise
forcing, the transport coefficients are approximately given
by~\cite{GM01}
\beq
\eta=\frac{p}{ \nu}\frac{4}{(1+\alpha)(3-\alpha)}\;, \quad
\kappa'=\frac{15p}{4m\nu }\frac{32}{(1+\alpha)(49-33\alpha)}\;.
\label{B1}
\eeq
While $\eta$ monotonically increases with inelasticity, $\kappa'$
starts decreasing with increasing inelasticity, reaches a minimum
value around $\alpha\simeq 0.4$, and then slightly increases for
$\alpha\lesssim 0.4$. The expressions for $C_T(\alpha)$,
$C_\varpi(\alpha)$, and $C_\varpi'(\alpha)$ are
\beq
C_T(\alpha)=\frac{1}{25}\frac{38+43\z+17{\z}^2}{(1+\z)(2+\z)}\;,\quad
\z\equiv\frac{\frac{5}{12}(1-\alpha^2)}{\beta(\alpha)+\frac{5}{12}(1-\alpha^2)}\;,
\eeq
\beq
C_\varpi(\alpha)=\frac{16}{25}\frac{82+67\z+8{\z}^2}{(1+\z)(2+\z)^2}\;,\quad
C_\varpi'(\alpha)=\frac{12}{25}\frac{102+87\z+13{\z}^2}{(1+\z)(2+\z)^2}\;.
\eeq
Of course, in the elastic limit ($\alpha\to 1$) the above
coefficients reduce to the BGK values listed in Table \ref{table1}.
The coefficient $C_T(\alpha)$ monotonically decreases with
increasing inelasticity. However, the non-monotonic behavior of
$\kappa'(\alpha)$ makes the relative location of the maximum
$y_{\text{max}}/\lambda_0$, where $\lambda_0=(\sqrt{2}\pi
n_0\sigma^2)^{-1}$, as well as the relative height
$(T_{\text{max}}-T_0)/T_0$ for a given value of $g\lambda_0/v_0^2$,
exhibit a non-monotonic behavior. This is illustrated in Fig.\
\ref{fig2}.

\section{Conclusions\label{sec5}}
In this paper we have reviewed the main results derived within a
kinetic theory  framework  for the steady Poiseuille flow driven by
a uniform longitudinal force. The results are restricted to the bulk
domain and are obtained  by a perturbation expansion through second
order. We have considered both elastic and inelastic particles. In
the first case, the planar and the cylindrical geometries have been
considered and the results are presented from the Boltzmann equation
for Maxwell molecules as well as from the BGK kinetic model for any
interaction. In the case of granular gases (inelastic collisions),
however, only the planar geometry and a BGK-like kinetic model have
been addressed.

In all the cases, the  profiles of the hydrodynamic fields and their
fluxes exhibit important deviations from the NS predictions: the
hydrostatic pressure is not uniform, normal stress differences are
present, a component of the heat flux normal to the thermal gradient
exists, and  the temperature profile  has a non-monotonic shape. In
general, those deviations are more important in the planar geometry
than in the cylindrical one. For instance, the  relative height of
the maximum temperature, $(T_{\text{max}}-T_0)/T_0$, is about five
times larger in the former case than in the latter, even though the
location of that maximum is similar in both cases, i.e.,
$y_{\text{max}}\simeq r_{\text{max}}$. It is worthwhile noting that
the BGK model succeeds in capturing the functional form of the
profiles derived from a more fundamental Boltzmann description,
albeit with some changes in the numerical factors. Thus, the BGK
model underestimates the maximum value $T_{\text{max}}$, as well as
its location $y_{\text{max}}$ or $r_{\text{max}}$. Concerning the
case of inelastic collisions, the results show a weak influence of
the coefficient of restitution $\alpha$. In any case, for small and
 moderate inelasticities (say $\alpha\gtrsim 0.5$) there is a slight decrease in the
quantitative deviations from the NS profiles as inelasticity grows,
while the opposite behavior takes place for high inelasticity
($\alpha\lesssim 0.5$).

\end{document}